\documentclass[prd,a4paper,showpacs,showkeys,preprint,byrevtex]{revtex4}
\usepackage{graphicx}
\usepackage{dcolumn}
\usepackage{amsmath}
\usepackage{bm}

\begin{document}

\title{Charm and bottom baryon masses in the combined $1/N_c$ and 
$1/m_Q$ expansion versus quark model}
\author{Claude \surname{Semay}}
\email[E-mail: ]{claude.semay@umh.ac.be}
\author{Fabien \surname{Buisseret}}
\email[E-mail: ]{fabien.buisseret@umh.ac.be}
\affiliation{Groupe de   
Physique Nucl\'{e}aire Th\'{e}orique,
Universit\'{e} de Mons-Hainaut,
Acad\'{e}mie universitaire Wallonie-Bruxelles,
Place du Parc 20, B-7000 Mons,  
 Belgium.}
\author{Florica \surname{Stancu}}
\email[E-mail: ]{fstancu@ulg.ac.be}
\affiliation{University of Li\`ege, Institute of Physics B5, Sart Tilman,
B-4000 Li\`ege 1, Belgium.}

\date{\today}

\begin{abstract}
A good agreement between a flux tube-based quark model of light baryons 
(strange and nonstrange) and the $1/N_c$ expansion mass formula has been 
found in previous studies. In the present work a larger connection is
established  between the quark 
model and the $1/N_c$ and $1/m_Q$ expansion method by extending the previous procedure
to baryons made of one heavy and two light quarks. The compatibility between both approaches is shown to hold in this sector too.  
\end{abstract}




\maketitle

\section{Introduction}

The recent discoveries of the $\Xi_b$, $\Sigma_b$ and $\Sigma^*_b$ baryons
at the Tevatron have incited to a new analysis of heavy baryons 
both within the combined $1/N_c$ and $1/m_Q$ expansion \cite{Jenkins:2007dm}
and the quark model, see for example Refs.~\cite{Roberts:2007ni,Karliner:2006ny,Karliner:2008sv}. 
The combined $1/N_c$ and $1/m_Q$ expansion is a model independent method.
It is thus important to search for a link between this method  
and the quark model. In previous studies \cite{lnc,lnc2} we have investigated
the possibility to establish a connection between the two approaches and 
we have found that a remarkable compatibility exists between them
when dealing with nonstrange \cite{lnc} or strange baryons \cite{lnc2}.

Presently we extend the ideas of our previous studies \cite{lnc,lnc2} to the 
case of heavy baryons made of one heavy quark ($c$ or $b$) and two light ones 
($u$, $d$, or $s$). This is the first step of a larger project and we view it
as an exploratory work where we search for the compatibility between the
spin-independent part of a quark model Hamiltonian and the corresponding
terms in the combined  $1/N_c$ and $1/m_Q$ expansion mass formula for the
ground state.
The spin-dependent part as well as the excited states
will be analyzed subsequently.

As previously, the comparison of the quark model results with those of
the $1/N_c$ expansion, presently combined with an  $1/m_Q$ expansion,
will be based on the introduction of a quantum number $N$, which is the 
same as in the harmonic oscillator potential and which 
is treated as a band number in baryon phenomenology. The introduction 
of $N$ in the eigenvalues
of the Hamiltonian was quite simple for identical quarks, the procedure
becomes more involved for baryons containing heavy quarks, as we shall see.  
 
The paper is organized as follows. After a summary of the charm and bottom 
baryon flavor states given in Sec.~\ref{FS}, the mass formula used by 
combining the $1/N_c$ and $1/m_Q$ expansions for such baryons is presented 
in Sec.~\ref{h_bar}. Section~\ref{qmhb}  gives a corresponding mass 
formula obtained from a Hamiltonian quark model where the confinement is 
of Y-junction type and where one gluon exchange and quark self-energy 
contributions are added perturbatively. 
In that section the excitation quantum number $N$ is introduced and its meaning 
is discussed.
A comparison between  results 
obtained on one hand in the combined $1/m_Q$ and $1/N_c$ expansion and 
on the other hand in the quark model  is then made 
in Sec.~\ref{compar}. Conclusions are finally drawn in Sec.~\ref{conclu}.

In the following, the symbol $q$ will denote a light quark ($u$, $d$, $s$)
and the symbol $Q$ will denote a heavy quark ($c$, $b$). Moreover, the symbol 
$n$ will be used for $u$ and $d$ quarks since both particles are assumed to 
have the same mass, as in our previous works. 

\section{Flavor states}\label{FS}

\subsection{Charm baryons}

Here we introduce the classification of ground state heavy baryons based 
on SU(4). 
In the following the total spin of a baryon is denoted by $\vec J$, the 
spin of the light subsystem by $\vec J_{qq}$, and that of the heavy quark by 
$\vec J_Q$.
In SU(4) the baryon multiplets arise from the direct product decomposition
${\bm 4} \otimes\bm{ 4} \otimes\bm{ 4} = \bm{20} \oplus \bm{20}\oplus \bm{20}\oplus\bm{4}$, 
see e.~g. Ref.~\cite{book}.
All baryons in the symmetric multiplet $ {\bf 20} $ have 
$J^P = {\frac{3}{2}}^+$.  The lightest SU(3) submultiplet is 
the well known Gell-Mann-Ne'eman decuplet. The single charm    
baryons form a sextet where the Fermi statistics requires
 $J_{qq}$ = 1. The six baryons 
$\Sigma^{*++}_c$,  $\Sigma^{*+}_c$, $\Sigma^{*0}_c$, $\Xi^{*'+}_c$,
$\Xi^{*'0}_c$ and  $\Omega^{*0}_c$
have the flavor structure
given in Table~\ref{symflstates}.
The remaining members of
the symmetric multiplet are the three double charm baryons
$\Xi^{*+}_{cc}$, $\Xi^{*++}_{cc}$,
$\Omega^{*+}_{cc}$ and the triple charm baryon $\Omega^{+++}_{ccc}$.

\begin{table}[tbp]
\centering
\caption{Flavor states of the single charm sextet baryons: $J^P =
{\frac{1}{2}}^+$ (no star) for baryons in the mixed representation and
 $J^P =  {\frac{3}{2}}^+ $ (with star) for baryons in the symmetric 
 representation.  They all have $J_{qq}$ = 1. Members of the same doublet
 become degenerate at $m_Q \rightarrow \infty$.\label{symflstates}}
\begin{ruledtabular}
\begin{tabular}{cr}
Baryon doublet &   Flavor state \\
\hline
$\Sigma^{++}_c$, $\Sigma^{*++}_c$ & $uuc$ \\
$\Sigma^{+}_c$, $\Sigma^{*+}_c$ & $\frac{1}{\sqrt{2}}\left( ud+du \right)c$ \\
$\Sigma^{0}_c$, $\Sigma^{*0}_c$     & $ddc$ \\
$\Xi^{'+}_c$, $\Xi^{*'+}_c$  & $\frac{1}{\sqrt{2}}\left( us+su \right)c $ \\
$\Xi^{'0}_c$, $\Xi^{*'0}_c$ & $\frac{1}{\sqrt{2}}\left( ds+sd \right)c$\\
$\Omega^{0}_c$, $\Omega^{*0}_c$ & $ssc$\\
\end{tabular}
\end{ruledtabular}
\end{table}

The experimental masses of single charm baryons with $J^P = {\frac{3}{2}}^+$ are 
\cite{PDG}
\begin{eqnarray}
\Sigma^*_c  & = & 2518.0  \pm 0.8  ~ {\rm MeV}, \nonumber \\
\Xi^*_c     & = & 2646.4  \pm 0.9  ~ {\rm MeV}, \nonumber \\
\Omega^*_c  & = & 2768.3  \pm 3.0  ~ {\rm MeV}. 
\end{eqnarray}
which represent mass averages when the hadron appears with different charges.
Note that here and below none of the quantum numbers assigned to the charm
baryons have been measured experimentally, but are based on quark model 
expectations.

The mixed symmetric representation ${\bf 20}$ has $J^P = {\frac{1}{2}}^+$.
The lowest submultiplet is 
the SU(3) Gell-Mann-Ne'eman octet. The single charm baryons
$\Sigma^{++}_c$,  $\Sigma^{+}_c$, $\Sigma^{0}_c$, $\Xi^{'+}_c$,
$\Xi^{'0}_c$, and $\Omega^{0}$ form 
a sextet with flavor states indicated in Table~\ref{symflstates}
and $J_{qq}$ = 1. $\Lambda^{+}_c$, $\Xi^{+}_c$, and $\Xi^{0}_c$ form
an antitriplet with flavor states shown in 
Table~\ref{symflstates_antitrip} having $J_{qq}$ = 0.

\begin{table}[tbp]
\caption{Flavor states of the single charm antitriplet baryons with $J^P =
{\frac{1}{2}}^+$ in the mixed symmetric representation. They all have 
$J_{qq}$ = 0. \label{symflstates_antitrip}}
\begin{ruledtabular}
\begin{tabular}{cc}
Baryon  &   Flavor state \\
\hline
$\Lambda^{+}_c$  & $ \sqrt{\frac{1}{2}}~(ud-du)~c $ \\
$\Xi^{+}_c$      & $\sqrt{\frac{1}{2}}~(us-su)~c $ \\
$\Xi^{0}_c$        & $\sqrt{ \frac{1}{2}}~(ds-sd)~c $ \\
\end{tabular}  
\end{ruledtabular}
\end{table}

The experimental masses of single charm baryon with $J^P = {\frac{1}{2}}^+$
are \cite{PDG}
\begin{eqnarray}
\Lambda_c   & = & 2286.46 \pm 0.14 ~ {\rm MeV}, \nonumber \\
\Sigma_c    & = & 2453.56 \pm 0.16 ~ {\rm MeV}, \nonumber \\
\Xi_c       & = & 2469.5  \pm 0.3  ~ {\rm MeV}, \nonumber \\
\Xi'_c      & = & 2576.9  \pm 2.1  ~ {\rm MeV}, \nonumber \\
\Omega_c    & = & 2697.5  \pm 2.6  ~ {\rm MeV},
\end{eqnarray}
where, again, mass averages are made when the hadron appears with different 
charges. In the observed $\Xi_c$  and $\Xi'_c$ baryons it is
expected that the light quarks are mostly in a state with $J_{qq}$ = 0 and 
$J_{qq}$ = 1 respectively.   

The mixed symmetric multiplet also contains three double charm baryons 
$\Xi^+_{cc}$, $\Xi^{++}_{cc}$, and $\Omega^+_{cc}$ from which only  
$\Xi^+_{cc}$ has been observed  by SELEX with a mass 
of 3518.9$\pm$0.9 MeV \cite{PDG}, but needs confirmation.


\subsection{Bottom baryons}

Despite the large symmetry breaking, for the sake of the 
classification one can also assume
an SU(4) classification of bottom baryons.
Similarly, for single bottom baryons there is a sextet
shown in Table~\ref{b_sym}
and an antitriplet shown in Table~\ref{b_antitrip}.
The mass of $\Lambda_b$ has been previously measured \cite{PDG}
\begin{eqnarray}
\Lambda_b   & = & 5620.2 \pm 1.6 ~ {\rm MeV}.
\end{eqnarray}   

\begin{table}[tbp]
\caption{Flavor states of the single bottom sextet baryons: $J^P =
{\frac{1}{2}}^+$ (no star) for baryons in the mixed representation and
 $J^P =  {\frac{3}{2}}^+ $ (with star) for baryons in the symmetric 
 representation.  They all have $J_{qq}$ = 1. Members of the same doublet
 become degenerate at $m_Q \rightarrow \infty$. \label{b_sym}}
\begin{ruledtabular}
\begin{tabular}{cr}
Baryon doublet &   Flavor state \\
\hline
$\Sigma^{+}_b$, $\Sigma^{*+}_b$ & $uub$ \\
$\Sigma^{0}_b$, $\Sigma^{*0}_b$ & $\frac{1}{\sqrt{2}}\left( ud+du \right) b$ \\
$\Sigma^{-}_b$, $\Sigma^{*-}_b$ & $ddb$ \\
$\Xi^{'0}_b  $, $\Xi^{*'0}_b  $ & $\frac{1}{\sqrt{2}}\left( us+su \right) b $ \\
$\Xi^{'-}_b  $, $\Xi^{*'-}_b  $ & $\frac{1}{\sqrt{2}}\left( ds+sd \right) b $\\
$\Omega^{-}_b$, $\Omega^{*-}_b$ & $ssb$\\
\end{tabular}
\end{ruledtabular}
\end{table}

\begin{table}[tbp]
\caption{Flavor states of the single bottom antitriplet baryons with $J^P =
{\frac{1}{2}}^+$ in the mixed symmetric representation. They all have 
$J_{qq}$ = 0. \label{b_antitrip}}
\begin{ruledtabular}
\begin{tabular}{cr}
Baryon  &   Flavor state \\
\hline
$\Lambda^{+}_b$  & $ \sqrt{\frac{1}{2}}~(ud-du)~b $ \\
$\Xi^{+}_b$      & $\sqrt{\frac{1}{2}}~(us-su)~b $ \\
$\Xi^{0}_b$        & $\sqrt{ \frac{1}{2}}~(ds-sd)~b $ \\
\end{tabular}
\end{ruledtabular}
\end{table}

Recent measurements have been made for $\Xi_b$ \cite{:2007ub,CDF1},
$\Sigma_b$, and $\Sigma^*_b$ \cite{CDF2}. The measured masses are
\begin{alignat}{1}
\Xi^{-}_b & =  5774 \pm 11 \pm 15~{\rm MeV}\ \text{\cite{:2007ub}},  
\ 5792.9 \pm 2.5 \pm 1.7~{\rm MeV}\  \text{\cite{CDF1}}, \nonumber \\
\Sigma^{\pm}_b & =  5811.5 \pm 1.7~{\rm MeV}\ \text{\cite{CDF2}}, \nonumber \\
\Sigma^{*\pm}_b & =  5832.7 \pm 1.8~{\rm MeV}\ \text{\cite{CDF2}}. 
\end{alignat}
The remaining undiscovered single bottom baryons are
$\Xi^{'}_b$, $\Xi^{*'}_b$, $\Omega_b$, and $\Omega^*_b$. 
 
\section{Ground state heavy baryons in the $1/m_Q$ and $1/N_c$ expansion}\label{h_bar}

The approximate spin-flavor symmetry for large $N_c$ baryons 
containing light $q = u, d$, or $s$ quarks and heavy $Q = c$ or $b$ quarks
is SU(6)$\times$ SU(2)$_c$ $\times$  SU(2)$_b$,
\emph{i.e.} there is a separate spin symmetry for each heavy flavor.
Over a decade ago the $1/N_c$ expansion has been
generalized to include an expansion   
in $1/m_Q$ and light quark flavor 
symmetry breaking \cite{Jenkins:1996de}. 

Let us first consider that SU(3)-flavor symmetry is exact. In 
this case the mass operator is a flavor singlet.  
In the combined $1/m_Q$ and $1/N_c$ expansion to order $1/m_Q^2$ 
the ground state mass operator $M^{(1)}$ takes the following 
form
\begin{equation}\label{mlnc}
M^{(1)}=m_Q N_Q  \openone
+\Lambda_{qq}+\lambda_Q+\lambda_{qqQ},
\end{equation}
where $N_Q$ is the number of heavy quarks. 
The leading order term 
is $m_Q$ at all orders in the $1/N_c$ expansion. Next we have 
\begin{equation}
\Lambda_{qq}=c_0\, N_c\, \openone+\frac{c_2}{N_c}\, J^2_{qq},
\end{equation}
where $\vec J_{qq}$ is the total spin of the light quark pair. 
This operator contains the dynamical contribution of the light 
quarks and is independent of $m_Q$.  
Then, $\lambda_Q$ gives the $1/m_Q$ corrections due to $N_Q$ heavy quarks
\begin{equation}
\lambda_Q=N_Q \frac{1}{2 m_Q} \left(c^{'}_0\, 
\openone+\frac{c^{'}_2}{ N^2_c}J^2_{qq}\right).  
\end{equation}
In the following we shall deal with $N_Q$ = 1 only.
Lastly, $\lambda_{qqQ}$ contains the 
heavy-quark spin-symmetry violating (chromomagnetic) operator which is
of order $1/m_Q$ as well
\begin{equation}
\lambda_{qqQ}=2 \frac{c^{''}_2}{N_c m_Q}\vec J_{qq}\cdot \vec J_Q,
\end{equation}
$\vec J_Q$ being the  spin of the heavy quark. This is the term responsible
for the splitting between states which form  degenerate doublets in the heavy 
quark limit, see Tables~\ref{symflstates} and \ref{b_sym}.

The unknown coefficients $c_0$, $c_2$, $c^{'}_0$, $c^{'}_2$, and  $c^{''}_2$
are functions of $1/N_c$ and of a QCD scale $\Lambda$.
Each coefficient has an expansion in $1/N_c$ where the leading term is 
of order unity and does not depend on $1/m_Q$.
Without loss of generality one can set $c_0 \equiv \Lambda$.
The other coefficients contain a dimensional power of $\Lambda$ and
a dimensionless function of $1/N_c$ beginning at order unity 
and  have to be fitted to the available experimental data. 
In agreement with Ref.~\cite{Jenkins:1996de}, we can take
\begin{align}
\label{largenpar}
&c_0 =  \Lambda, c_2 \sim \Lambda,  \nonumber \\ 
&c^{'}_0  \sim   c^{'}_2\sim c^{''}_2  \sim  {\Lambda}^2.   
\end{align}

\begin{table}[tbp]
\caption{Mass combinations resulting from heavy quark and large $N_c$
limit and their experimental values \cite{Jenkins:2007dm}. \label{masscomb}}
\begin{ruledtabular}
\begin{tabular}{lcc}
Mass combination  &   Exp  (MeV)   &  Exp  (MeV)\\
                  &   $Q = c$ &  $Q = b$ \\
\hline
$\Lambda_Q$  &  2286.46$\pm$0.14  &  5620.2$\pm$1.6 \\
$\frac{1}{3}(\Sigma_Q + 2 \Sigma^{*}_Q)-\Lambda_Q$ & 210.0$\pm$0.5 & 205.4$\pm$2.1\\
$\Sigma^{*}_Q - \Sigma_Q$                          & 64.4$\pm$0.8  &  21.2$\pm$2.5\\
$	\Xi_Q-\Lambda_Q$ & 183.0$\pm$0.3 & 172.7$\pm$3.4\\
\end{tabular}
\end{ruledtabular}
\end{table}

At the dominant order, the value of $\Lambda$ can be extracted from the mass combinations 
\begin{subequations}\label{mcomb}
\begin{equation}\label{E1}
\Lambda_Q= m_Q+ N_c \Lambda,
\end{equation}
\begin{equation}\label{E2}
\frac{1}{3} (\Sigma_Q + 2 \Sigma^{*}_Q) - \Lambda_Q
= 2 \frac{\Lambda}{N_c},
\end{equation}  
\begin{equation}\label{E3}
\Sigma^{*}_Q - \Sigma_Q =  \frac{3}{2} \left(\frac{2 \Lambda^2}{N_c m_Q}\right),
\end{equation}
\end{subequations}
resulting from the mass definition (\ref{mlnc}). The equations (\ref{E1}) 
and (\ref{E2}) express the fact that $\lambda_Q$ is negligible with
respect to the other terms in (\ref{mlnc}). 
Here and below the particle label represents its mass.

A slightly more complicated 
mass combination, involving light baryons as well as heavy ones, directly 
leads to $m_Q$, that is \cite{Jenkins:2007dm}
\begin{equation}\label{E0}
\frac{1}{3}(\Lambda_Q+2\Xi_Q)-\frac{1}{4}\left[\frac{5}{8}(2N+3\Sigma+\Lambda+2\Xi)-\frac{1}{10}(4\Delta+3\Sigma^*+2\Xi^*+\Omega)\right]= m_Q.
\end{equation}
This mass combination gives 
\begin{subequations}\label{paramLNC}
\begin{equation}\label{mQdef}
	m_c=1315.1\pm0.2~{\rm MeV},\quad m_b=4641.9\pm2.1~{\rm MeV},
\end{equation}
while the value 
\begin{equation}\label{lambdadef}
	\Lambda \approx 324~{\rm MeV}
\end{equation}
\end{subequations}
ensures that the mass combinations~(\ref{mcomb}) are optimally compatible with
the experimental values for $Q = c$ and $Q = b$ indicated in 
Table~\ref{masscomb}. Note also that the heavy quark flavor symmetry
predicts that the observed $(\Lambda_b - \Lambda_c) = 3333.7 \pm 1.6$ MeV
splitting \cite{Jenkins:2007dm}
can give a measure of the quark mass difference $m_b - m_c$
up to corrections of the order $\Lambda^2(1/2m_c-1/2m_b)\approx 23$~MeV
\cite{Jenkins:1996de}. The values given by Eqs.~(\ref{paramLNC})
satisfy this constraint.

The operator analysis including SU(3)-flavor breaking leads to an 
expansion in the SU(3) violating parameter $\epsilon$ which contains 
the singlet $M^{(1)}$, an octet $M^{(8)}$, and a 27-plet $M^{(27)}$.
The last term brings contributions proportional to $\epsilon^2$
and we neglect it. For $M^{(8)}$ we retain its dominant contribution 
$T^{8}$ to order $N^0_c$. Then the mass formula becomes
\begin{equation}
\label{breaksim} 
M = M^{(1)} + \epsilon T^{8}.
\end{equation}
The flavor breaking parameter $\epsilon$ is governed by the mass 
difference $m_s - m$ (where $m$ is the average of the $m_u$ and $m_d$ masses) 
and therefore is $\epsilon \sim 0.2$-0.3. It is measured in units of
the chiral symmetry breaking scale parameter $\Lambda_{\chi} \sim 1$ GeV.
A measure of the SU(3)-flavor breaking factor can be given by \cite{Jenkins:1996de}
\begin{equation}\label{break}
	\Xi_Q-\Lambda_Q=  \frac{\sqrt 3}{2}\, (\epsilon\Lambda_\chi).
\end{equation}
The value $(\epsilon\Lambda_\chi)=206$~MeV leads 
to $\Xi_Q-\Lambda_Q=178$~MeV, which is the average value of the 
corresponding experimental data listed in Table~\ref{masscomb}. 

\section{Quark model for heavy baryons}
\label{qmhb}

\subsection{Hamiltonian}  

The potential model used to describe heavy baryons is the same as that which
has been proposed in Ref.~\cite{lnc2} for light baryons. Let us recall 
its main features. 

In quark models, a baryon is a bound state of three valence quarks which can 
be described at the dominant order by the spinless Salpeter Hamiltonian
\begin{equation}
\label{ssh}
H=\sum^3_{i=1}\sqrt{\vec p^{\, 2}_i+m^2_i}+V_Y,
\end{equation}
where $m_i$ is the current (bare) mass of the quark $i$ and $V_Y$ is the confining 
interaction potential. Both the flux tube model \cite{CKP} and lattice 
QCD \cite{Koma,Okiharu} suggest that the flux tubes form a Y-junction: A flux tube 
starts from each quark and the three tubes meet at the Torricelli point of the 
triangle formed by the three quarks. This point, located in $\vec x_{T}$, minimizes the sum 
of the flux tube lengths, leading to the following confining potential 
\begin{equation}
\label{vYdef}
V_Y=a \sum^{3}_{i=1} \left|\vec{x}_{i}-\vec{x}_{T}\right|.
\end{equation}
The position of the quark $i$ is denoted by $\vec x_i$, and $a$ is the energy 
density of the flux tubes. Such a Hamiltonian can also be obtained in the framework
of the field correlator method \cite{dosc87}.

As $\vec x_T$ is a complicated three-body function, 
it is interesting to approximate the confining potential by a more tractable 
form. In the following, we will use 
\begin{align}
\label{ssh2}
H_R&=\sum^3_{i=1}\sqrt{\vec p^{\, 2}_i+m^2_i}+V_R, \\
\label{pot1}
V_R&=k\,a \sum^{3}_{i=1}\left|\vec{x}_{i}-\vec{R}\right|,
\end{align}
where $\vec{R}$ is the position of the center of mass and  $k$ 
is a corrective factor \cite{Bsb04}. The eigenvalues corresponding 
to potentials $V_Y$ and $V_R$ differ from each other only by about 5\% 
in most cases. The accuracy of the formula~(\ref{pot1}) is thus rather 
satisfactory, and has already led to relevant results in Ref.~\cite{lnc2}. 
For light (symmetrical) $qqq$ baryons, a good value for the corrective factor is $k_0=0.952$. 
For very asymmetrical $qqQ$ baryons, a good choice is 
$k_1=0.930$ \cite{Bsb04}. This last value corresponds actually to the case 
$m_q/m_Q \rightarrow 0$. 

Besides the confining potential (\ref{vYdef}),
other contributions are necessary to reproduce the baryon masses. We 
shall add them as perturbations to the dominant Hamiltonian~(\ref{ssh2}). 
The most widespread correction is a Coulomb interaction term of the form
\begin{equation}
\Delta H_{oge}=-\frac{2}{3}\sum_{i<j}\frac{\alpha_{S,ij}}{|
\vec x_i-\vec x_j|},
\end{equation}
arising from one gluon exchange processes, where $\alpha_{S,ij}$ is the
strong coupling constant between the quarks $i$ and $j$. Actually, one should deal 
with a running form $\alpha_S(r)$, but it would considerably increase the 
difficulty of the computations. Typically, we need two values: 
$\alpha_0=\alpha_{S,qq}$ for a $qq$ pair and $\alpha_1=\alpha_{S,qQ}$ for a 
$qQ$ pair, in the spirit of what has been done in a previous study describing 
mesons in the relativistic flux tube model \cite{tf_Semay}. There it was found that $\alpha_1/\alpha_0 \approx 0.7$ describes rather well the 
experimental data of $q\bar q$ and $Q\bar q$ mesons.   

Another perturbative contribution to the  
 mass is the quark self-energy. This is 
due to the color magnetic moment of a quark propagating through the 
vacuum background field. It adds a negative contribution to the hadron masses \cite{qse}. The quark self-energy contribution for a baryon is given by
\begin{equation}
\label{qsedef}
\Delta H_{qse}=-\frac{fa}{2\pi}\sum_i\frac{\eta(m_i/\delta)}{\mu_i}.
\end{equation}
The factors $f$ and $\delta$ have been computed in quenched and unquenched 
lattice QCD studies \cite{qse2,qse3}. Although it is not known with great 
accuracy, it seems well established that 
$3 \leq f\leq 4$ and ($1.0 \leq \delta \leq 1.3$)~GeV \cite{qse2,qse3}. 
The function $\eta(\epsilon)$ is analytically known; we refer the reader to 
Ref.~\cite{qse} for the explicit formula. For typical values of the light 
quark masses, we have $0 \leq m_q/\delta \lesssim 0.3$, while for heavy quarks, we have $1.0 \lesssim m_Q/\delta \lesssim 6.0$. 
The function $\eta(\epsilon)$ is such that
\begin{alignat}{2}
\label{etaap0}
\eta(\epsilon)&\approx 1+\left( 4+3\ln \frac{\epsilon}{2} \right) \epsilon^2 
&& \quad\textrm{for} \quad \epsilon \ll 1,\nonumber \\
&\approx\frac{2}{\epsilon^2} && \quad\textrm{for} \quad \epsilon\rightarrow\infty.
\end{alignat}
For the relevant values of $\epsilon = m_i/\delta$ a better accuracy is obtained with the following simple forms
\begin{alignat}{3}
\label{etaap}
\eta(\epsilon)&\approx 1-\beta \epsilon^2 && \quad\textrm{with} \quad \beta=2.85
&& \quad\textrm{for} \quad 0 \le \epsilon \lesssim 0.3,\nonumber \\
&\approx \frac{\gamma}{\epsilon^2} && \quad\textrm{with} \quad \gamma=0.79
&& \quad\textrm{for} \quad 1.0 \lesssim \epsilon \lesssim 6.0. 
\end{alignat}
Let us note that the corrections depending on the parameter $\gamma$ appear at order $1/m_Q^3$ in the mass formula, so they are not considered in this work.
Finally, $\mu_i$ is the dynamical mass of the quark $i$, defined as \cite{qse} 
\begin{equation}\label{muidef}
\mu_i=\left\langle \sqrt{\vec p^{\, 2}_i+m^2_i}\right\rangle.
\end{equation}
This dynamical mass is state-dependent: It represents the kinetic energy of 
the quark $i$ averaged with the wave function of the unperturbed spinless 
Salpeter Hamiltonian~(\ref{ssh2}).

\subsection{General formulas}

We are mainly interested in analytical expressions, so that a comparison with 
the large $N_c$ mass formula will be straightforward. To this aim, the auxiliary field technique will be used in order to transform the Hamiltonian~(\ref{ssh2}) into an analytically solvable one \cite{Sem03,silv08}. With $\lambda=k\, a$, we obtain
\begin{eqnarray}\label{ham3b}
H(\mu_i,\nu_j)=\sum^3_{j=1}\left[\frac{\vec{p}^{\, 2}_j+m
^2_j}{2\mu_j}+\frac{\mu_j}{2}\right]+\sum^3_{j=1}\left[\frac{ \lambda^2 (\vec{x}_j-\vec{R})^2}{2\nu_j}+
\frac{\nu_j}{2}\right].
\end{eqnarray}
The auxiliary fields, denoted as $\mu_i$ and $\nu_j$, are operators, and $H(\mu_i,\nu_j)$ is equivalent to $H$ up to their elimination thanks to the constraints
\begin{align}
\label{elim}
\delta_{\mu_i}H(\mu_i,\nu_j)&=0\ \Rightarrow\ \mu_{i,0}=
\sqrt{\vec{p}^{\, 2}_i+m^2_i}, \nonumber \\
\delta_{\nu_j}H(\mu_i,\nu_j)&=0\ \Rightarrow\ \nu_{i,0}=
\lambda|\vec{x}_i-\vec{R}|.
\end{align}
$\left\langle \mu_{i,0}\right\rangle$ is the dynamical quark mass introduced 
in Eq.~(\ref{muidef}), and $\left\langle \nu_{i,0}\right\rangle$ is the energy of the flux tube linking the quark $i$ to the center of mass. 

Although the auxiliary fields are operators, the calculations are considerably 
simplified if one considers them as real numbers. They are finally eliminated 
by a minimization of the masses \cite{Sem03}, and the extremal values of 
$\mu_i$ and $\nu_j$ are logically close to 
$\left\langle \mu_{i,0}\right\rangle$ and 
$\left\langle \nu_{j,0}\right\rangle$ respectively. This technique can give 
approximate results very close to the exact ones (see Ref.~\cite{naro08} for 
a comparative study of baryons with the auxiliary fields introduced only in 
the kinetic part of the Hamiltonian). 

In Ref.~\cite{coqm}, it has been shown that the eigenvalues of a Hamiltonian of the form (\ref{ham3b}) can be analytically found by
making an appropriate change of variables, the quark coordinates $\vec x_{i}=\left\{\vec x_{1},\vec x_{2},\vec x_{3}\right\}$ being
replaced by new coordinates $\vec x^{\, '}_{k}=\{\vec R,\vec \xi,\vec \eta\,\}$. The center
of mass is defined as \begin{equation}\label{cmdef}
\vec R=\frac{\mu_{1}\vec x_{1}+\mu_{2}\vec x_{2}+\mu_{3}\vec x_{3}}{\mu_{t}},
\end{equation}
with $\mu_{t}=\mu_{1}+\mu_{2}+\mu_{3}$. $\{\vec \xi,\vec \eta\,\}$ are 
two relative coordinates: $\vec \xi \propto \vec x_1-\vec x_2$ and 
$\vec \eta \propto \frac{\mu_1 \vec x_1+\mu_2 \vec x_2}{\mu_1+\mu_2}-\vec x_3$.
As we only consider baryons built from two different quarks, the general 
formulas obtained in Ref.~\cite{coqm} can be simplified. In the case of two 
quarks with mass $m$ and another with mass $m_3$, the mass spectrum of 
the Hamiltonian~(\ref{ham3b}) is given by ($\mu_1=\mu_2=\mu$, $\nu_1=\nu_2=\nu$)
\begin{equation}\label{mass1}	
M(\mu,\mu_3,\nu,\nu_3)=\omega_\xi(N_\xi+3  
/2)+\omega_\eta(N_\eta+3/2)+\mu+\nu+\frac{\mu_3+\nu_3}{2}+\frac{m^2}{\mu}+\frac{m^2_3}{2\mu_3},
\end{equation}
where
\begin{equation}
\omega_\xi=\frac{\lambda}{\sqrt{\mu\nu}},\quad \omega_\eta=\frac{\lambda}{\sqrt{2\mu+\mu_3}}\sqrt{\frac{\mu_3}{\mu\nu}+\frac{2 \mu}{\mu_3\nu_3}}.
\end{equation}
The integers $N_{\xi/\eta}$ are given by $2n_{\xi/\eta}+\ell_{\xi/\eta}$, 
where $n_{\xi/\eta}$ and $\ell_{\xi/\eta}$ are  the radial and 
orbital quantum numbers relative to the variable $\vec \xi/\vec \eta$
respectively. One can also easily check that \cite{coqm}
\begin{equation}
\left\langle \vec \xi^{\, 2}\right\rangle=\frac{N_\xi+3/2}{\phi\,\omega_\xi},\quad \left\langle \vec \eta^{\, 2}\right\rangle=\frac{N_\eta+3/2}{\phi\,\omega_\eta},
\end{equation}
with
\begin{equation}
\phi=\sqrt{\frac{\mu^2\mu_3}{2\mu+\mu_3}}.
\end{equation}
These last identities provide relevant informations about the structure of the baryons, since
\begin{align}
\left\langle \vec X^{\, 2}\right\rangle&=\left\langle (\vec x_1-\vec x_2)^{2}\right\rangle=\sqrt{\frac{4\mu_3}{2\mu+\mu_3}}\, \left\langle \vec\xi^{\, 2}\right\rangle,\label{X2def}\\
\left\langle \vec Y^{\, 2}\right\rangle&=\left\langle \left(\frac{\vec x_1+\vec x_2}{2}-\vec x_3\right)^2\right\rangle=\sqrt{\frac{2\mu+\mu_3}{4\mu_3}}\, \left\langle \vec\eta^{\, 2}\right\rangle.\label{Y2def}
\end{align}
Moreover, by symmetry, we can assume the following equality
\begin{equation}
\left\langle (\vec x_1-\vec x_3)^2\right\rangle	=\left\langle (\vec x_2-\vec x_3)^2\right\rangle\approx\frac{\left\langle \vec X^{\, 2}\right\rangle}{4}+  \left\langle \vec Y^{\, 2}\right\rangle,	
\end{equation}
which will be useful in the computation of the one gluon exchange contribution.

The case of $qqq$ baryons, studied in our previous papers \cite{lnc,lnc2}, 
is obtained by taking $m = m_n = 0$ and $m_3 = m_s$, 
and by setting 
$\lambda=k_0 a$. If the three quarks are identical, then $m_3=m$, $\mu_3=\mu$, 
$\nu_3=\nu$. For $qqQ$ baryons, we 
explicitly  write  $m_3=m_Q$, $\mu_3=\mu_Q$, $\nu_3=\nu_Q$, and we set 
$\lambda=k_1 a$, $m=0$ or $m_s$ for $n$ or $s$ quarks respectively.
Let us note that different values of $k_0$ have been previously 
used: $k_0=(1/2+\sqrt{3}/4)$ in Ref.~\cite{lnc} and $k_0=1$ in Ref.~\cite{lnc2}. 
In this work, we choose phenomenological values computed in Ref.~\cite{Bsb04} 
in order to obtain the best possible simulation of the Y-junction for both 
$qqq$ and $qqQ$ baryons with the potential~(\ref{pot1}). 

\subsection{Mass formula for heavy baryons}
\label{case1h}

In this section, we focus our attention on $ssQ$ baryons. The mass formula 
for $nnQ$ baryons is obtained simply by setting $m_s=0$, and the case of 
$nsQ$ baryons will be discussed in the next section. 
The four auxiliary fields appearing in the mass 
formula~(\ref{mass1}) have to be eliminated by solving simultaneously the four 
constraints:
\begin{alignat}{2}
\label{const}
&\partial_{\mu}M(\mu,\mu_Q,\nu,\nu_Q)=0, && \quad \partial_{\mu_Q}M(\mu,\mu_Q,\nu,\nu_Q)=0,\nonumber\\
&\partial_{\nu}M(\mu,\mu_Q,\nu,\nu_Q)=0, && \quad \partial_{\nu_Q}M(\mu,\mu_Q,\nu,\nu_Q)=0.
\end{alignat}
This cannot be done exactly in an analytical way, but solutions can be obtained by assuming that $1/m_Q$ and 
$m_s$ are small quantities. After some algebra, a solution was found by 
working at  order $1/m_Q$ and $m_s^2$ (all contributions proportional to $m_s$ are vanishing). By denoting
\begin{align}
\label{conv1}
N&=N_\xi+N_\eta, \nonumber \\
\mu_1 &= \sqrt{\frac{k_1 a (N+3)}{2}}, \nonumber \\
G(N,N_\eta)&=\sqrt{2 N_\eta+3}\left( \sqrt{2 (N+3)} - \sqrt{2 N_\eta+3} \right),
\end{align}
we have obtained
\begin{align}
\label{af1s}
\mu&=\mu_1 + \frac{3 m_s^2}{4 \mu_1} - \frac{k_1 a}{4 m_Q} G(N,N_\eta) ,\nonumber\\ 
\nu&=\mu_1 - \frac{m_s^2}{4 \mu_1} - \frac{k_1 a}{4 m_Q}(2N_\eta +3) ,\nonumber\\
\mu_Q&=m_Q + \frac{k_1 a}{2 m_Q} G(N,N_\eta) ,\nonumber\\ 
\nu_Q&=\frac{k_1 a}{m_Q}\sqrt{\frac{(2N_\eta +3)(N+3)}{2}} .
\end{align}
Logically, $\mu_Q\approx m_Q$ since this auxiliary field is dominated by the effective mass of the heavy quark. The length of the flux tube joining the heavy quark to the center of mass is smaller than the other ones, so $\lim_{m_Q \rightarrow \infty}\nu_Q =0$ as expected.    

The mass formula~(\ref{mass1}), in which the auxiliary fields are replaced by 
the expressions~(\ref{af1s}), reads at  orders $1/m_Q$ and $m_s^2$ as
\begin{equation}
\label{mass3}
M=m_Q+4\mu_1+\frac{m_s^2}{\mu_1}+\frac{k_1 a}{2m_Q} G(N,N_\eta).
\end{equation}
It is interesting to look at the magnitude of the various terms in this 
formula. Let us choose typical values for the parameters: $k_1=1$, 
$a=0.2$~GeV$^2$, $m_s=0.3$~GeV, $m_c=1.5$~GeV, $m_b=5.0$~GeV. For the ground 
state ($N=0$), $\mu_1=0.548$~GeV. The contribution of the kinetic energy and of
the confinement in $M$, given by $4 \mu_1=2.191$~GeV, is of the order of $m_Q$. The contribution of the strange quark is given by $\frac{m_s^2}{\mu_1}=0.164$~GeV, while the term $\frac{k_1 a}{2m_Q} G(0,0)$ is 0.083~GeV and 0.025~GeV respectively for the charm and bottom masses. These values justify \emph{a posteriori} the use of the power expansion in $m_s$ and in $1/m_Q$. 
Formulas~(\ref{af1s}) and (\ref{mass3}) giving the optimal values of the 
auxiliary fields and the corresponding minimal mass are approximate solutions 
of Eq.~(\ref{mass1}). In Table~\ref{tab:comp}, these values are compared with the exact solutions obtained numerically. In all cases, the error on the mass is quite small, even if the error on some auxiliary fields is larger. The auxiliary fields $\mu$ and $\mu_Q$ are used to compute perturbatively the self-energy. Fortunately, the error on these fields are small. As expected, the accuracy is improved for large values of $m_Q$, while $m_s$ has only a little influence. 

\begin{table}[htbp]
\centering
\caption{Relative error (\%) on auxiliary fields (\ref{af1s}) and mass (\ref{mass3}) for typical values of the physical parameters ($k_1=1$, $a=0.2$~GeV$^2$). Quark masses are given in GeV. \label{tab:comp}}
\begin{ruledtabular}
\begin{tabular}{lrrrrrrrrrrrr}
$m_s$ / $m_Q$ & \multicolumn{3}{c}{0 / 1.5} & \multicolumn{3}{c}{0.3 / 1.5} &
\multicolumn{3}{c}{0 / 5.0} & \multicolumn{3}{c}{0.3 / 5.0}\\
($N,N_\eta$) & (0,0) & (4,0) & (4,4) & (0,0) & (4,0) & (4,4)
& (0,0) & (4,0) & (4,4) & (0,0) & (4,0) & (4,4) \\
\hline 
$\mu$ & 0.007 & 6.6 & 5.2 & 2.8 & 5.6 & 5.4 & 0.2 & 0.7 & 1.4 & 2.9 & 0.02 & 2.0 \\
$\nu$ & 8.9 & 7.1 & 29.5 & 10.2 & 7.2 & 30.1 & 1.1 & 1.0 & 3.6 & 2.3 & 1.1 & 3.5 \\
$\mu_Q$ & 0.2 & 4.8 & 4.6 & 0.8 & 4.5 & 5.3 & 0.04 & 0.2 & 0.4 & 0.1 & 0.1 & 0.5 \\
$\nu_Q$ & 44.5 & 74.0 & 67.4 & 33.7 & 68.8 & 61.9 & 12.8 & 21.8 & 18.7 & 1.8 & 16.4 & 13.3 \\
$M$ & 0.2 & 1.4 & 1.7 & 0.5 & 1.7 & 1.7 & 0.05 & 0.1 & 0.3 & 0.6 & 0.3 & 0.2
\end{tabular}
\end{ruledtabular}
\end{table}

The contribution of the one gluon exchange term can be computed with the help 
of relations~(\ref{X2def}) and (\ref{Y2def}). One obtains 
\begin{align}
\Delta M_{oge}&\approx-\frac{2}{3} \left[\frac{\alpha_0}{\sqrt{\left\langle \vec X^{\, 2}\right\rangle}}
+\frac{2\alpha_1}{\sqrt{\left\langle \vec X^{\, 2}\right\rangle/4+  \left\langle \vec Y^{\, 2}\right\rangle}}\right]\nonumber\\
&=-\frac{2}{3}\alpha_0\sqrt{\frac{k_1 a}{2N_\xi+3}}\left[1+\frac{m_s^2}{4 \mu_1^2}+\frac{ \sqrt{k_1 a}}{8m_Q}\sqrt{2N_\eta+3}\left(\sqrt{\frac{2(2N_\eta+3)}{N+3}}-1\right)\right]\nonumber  \\
&\phantom{=}-\frac{4}{3}\alpha_1\sqrt{\frac{2k_1 a}{N+3}}\left[1+\frac{m_s^2}{4 \mu_1^2}-\frac{\sqrt{k_1 a}}{2m_Q}\frac{2N_\eta+3}{\sqrt{2(N+3)}}\right].
\label{oge1h}
\end{align}
For values of the parameters defined above together with 
$\alpha_0=\alpha_1=0.4$, 
the contribution of the dominant term in $\Delta M_{oge}$ is $-0.264$~GeV for 
the ground state. The $m_s^2/\mu_1^2$ term brings $-0.020$~GeV while the $1/m_c$ term brings $0.034$~GeV. Again, the use of the power expansion in $m_s$ and in $1/m_Q$ seems relevant.

The relations~(\ref{af1s}) defining $\mu$ and $\mu_Q$ allow to write down
the contribution of the quark self-energy~(\ref{qsedef}). Using the 
approximation~(\ref{etaap}) one obtains
\begin{equation}
\label{qse1s}
\Delta  M_{qse}=-\frac{f a}{\pi\mu_1}\left[1-\left( \frac{3}{4 \mu_1^2}+\frac{\beta}{\delta^2} \right) m_s^2+\frac{k_1 a}{4 \mu_1 m_Q} G(N,N_\eta)\right].
\end{equation}
We recall that the correction proportional to $\beta m_s^2$ comes from a convenient parameterization of the $\eta(\epsilon)$ function, while the term proportional to $m_s^2/\mu_1^2$ is due to the expansion of the auxiliary field $\mu$. 
For the values of the parameters defined above and the typical values $f=3.5$ and $\delta=1$~GeV, the contribution of the dominant term in $\Delta M_{qse}$ is $-0.302$~GeV for the ground state. The $m_s^2/\mu_1^2$ term brings $0.092$~GeV while the $1/m_c$ term brings $-0.031$~GeV. The use of the power expansion in $m_s$ and in $1/m_Q$ seems here more questionable, mostly for the contribution of the strange quark. This is due to the particular nature of the self-energy interaction which can be defined only as a perturbation \cite{qse}.

If we now look at the dominant terms in $M-m_Q$, $\Delta M_{oge}$ and $\Delta  M_{qse}$, we find respectively 2.191~GeV, $-0.264$~GeV and $-0.302$~GeV for the ground state with parameters defined above. These numbers show that it is \emph{a posteriori} justified to treat the Coulomb interaction and the self-energy interaction as perturbations.

\subsection{Mass formulas for general $qqq$ and $qqQ$ baryons}

In this section we gather mass formulas obtained for both light and heavy 
baryons. The $qqq$ mass formula is given in Ref.~\cite{lnc2} and is reminded 
here for completeness 
\begin{align}
\label{M_qqq}
\mu_0&=\sqrt{\frac{k_0 a(N+3)}{3}}, \nonumber \\
M_{qqq}&=M_0 + n_s\, \Delta M_{0s} \quad\quad (n_s=0,1,2,3), \nonumber \\
M_0 &= 6\mu_0-\frac{2 k_0 a \alpha_0}{\sqrt{3}\mu_0}-\frac{3 f a}{2\pi\mu_0}, \nonumber \\
\Delta M_{0s}&=\left[\frac{1}{2}-\frac{k_  
0 a \alpha_0}{6\sqrt{3} \mu_0^2}
+\frac{f a}{2\pi}\left(\frac{3}{4\mu^2_0}+
\frac{\beta}{\delta^2}\right) \right] \frac{m^2_s}{\mu_0}.
\end{align}
All parameters were already presented above, except the number $n_s$ of 
$s$-quarks in the baryons. The mass formula $M_{qqq}$ depends only on 
$N=N_\xi+N_\eta$ since the contribution of terms proportional to 
$N_\xi-N_\eta$, vanishing for $n_s=0$ and 3, was found to be 
very weak in general \cite{lnc2}.

In the previous section, only the case of a heavy baryon containing two 
identical light quarks was treated ($n_s=0$ or $n_s=2$). 
It has been shown that every $s$ quark brings the same contribution
$\Delta M_{0s}$ to the mass of a light baryon [see Eq.~(\ref{M_qqq})]. So, we 
can reasonably assume that the same situation occurs for $qqQ$ baryons. To take 
into account the contribution of $n_s$ quarks to the mass of these baryons, 
it is  enough to replace the term $m_s^2$ by $n_s m_s^2/2$ in Eqs.~(\ref{mass3}), 
(\ref{oge1h}) and (\ref{qse1s}).
Let us note that it is not necessarily true for the auxiliary fields $\mu$ and $\nu$ \cite{lnc2}. 
In the following formulas, we keep 
explicitly the dependence on both $N_\xi$ and $N_\eta$:
\begin{align}
\label{M_qqQ}
\mu_1&=\sqrt{\frac{k_1 a(N+3)}{2}}, \nonumber \\
M_{qqQ}&=m_Q+M_1 + n_s\, \Delta M_{1s} + \Delta M_{Q}\quad\quad (n_s=0,1,2), \nonumber \\
M_1 &=  4\mu_1 - \frac{2}{3}\left( \alpha_0\sqrt{\frac{k_1 a}{2N_\xi+3}} 
+ 2 \alpha_1\sqrt{\frac{2 k_1 a}{N+3}}\right)-\frac{f a}{\pi\mu_1}, \nonumber \\
\Delta M_{1s}&=\frac{m_s^2}{\mu_1} 
\left[ \frac{1}{2} 
- \frac{1}{12 \mu_1} \left( \alpha_0\sqrt{\frac{k_1 a}{2N_\xi+3}} 
+ 2 \alpha_1\sqrt{\frac{2 k_1 a}{N+3}}\right) 
+ \frac{f a}{2 \pi}\left( \frac{3}{4 \mu_1^2}+\frac{\beta}{\delta^2} \right) \right], \nonumber \\
\Delta M_{Q}&=\frac{k_1 a}{2m_Q} \left[ \left(1-\frac{f a}{2 \pi \mu_1^2}\right)G(N,N_\eta)
-\frac{\alpha_0}{6} \sqrt{\frac{2N_\eta+3}{2N_\xi+3}}\left(\sqrt{\frac{2(2N_\eta+3)}{N+3}}-1\right) 
\right. \nonumber \\
&\phantom{=\frac{k_1 a}{2m_Q} +}\left.+\frac{4 \alpha_1}{3} \frac{2N_\eta+3}{N+3} \right].
\end{align}

\subsection{What is the good quantum number?}

At the lowest order, the mass formula~(\ref{mass3}), with the rescaling $a \leftrightarrow \sigma$ (see next section), leads to
\begin{equation}
(M-m_Q)^2=\frac{4\pi\sigma}{3}\frac{k_1}{k_0}(N+3).
\end{equation}
The model thus predicts Regge trajectories for heavy baryons, with a slope of $4\pi\sigma k_1/(3 k_0) \approx 1.3 \pi \sigma$ instead of $2\pi\sigma$ for light baryons. At this dominant order, the mass formula depends only on $N$.
However, when corrections are added, the mass formula is no more symmetric in $N_\eta$ and $N_\xi$. Is it still possible to find a single quantum number? There are three possibilities:
\begin{itemize}
\item As in Ref.~\cite{lnc2}, we could assume that $N_\xi\approx N_\eta$. 
But, the presence of a heavy quark makes the system rather asymmetric 
in the $\vec \xi$ and $\vec \eta$ variables. So this solution seems unnatural.
\item Another possibility is to impose $N_{\eta}=0$ and $N_\xi=N$. With no 
excitation in the $\vec \eta$ variable, the two light quarks are moving around 
a static heavy quark in the configuration $q-Q-q$, as proposed in Ref.~\cite{hbar3}.
\item The opposite possibility can also be assumed: $N_\eta=N$ and $N_\xi=0$. 
With no excitation in the $\vec \xi$ variable, the two light quarks behave as a 
diquark orbiting around the heavy quark by forming a $Q-(qq)$ system, as 
considered in Ref.~\cite{hbar4}.
\end{itemize}

At order $1/m_Q$, the dominant term~(\ref{mass3}) depends on the function 
$G(N,N_\eta)$. The baryon mass is lowered when $G(N,N_\eta)$ is minimal, 
that is to say for $N_\eta=N$. In this case 
\begin{equation}
F(N) =G(N,N) = \sqrt{3+2N}\left[\sqrt{2(3+N)}-\sqrt{3+2N}\right],
\end{equation}
with $F(0)=3(\sqrt{2}-1) \leq F(N) < 3/2$, this upper bound being the limit 
of $F(N)$ for $N$ going to infinity. The analysis of the dominant part of the 
Coulomb term~(\ref{oge1h}) shows that the baryon mass is also lowered when 
$N_\eta=N$. So it is natural to assume that the favored configuration, 
minimizing the baryon energy, is $N_\eta=N$ and $N_\xi=0$, 
as in Ref.~\cite{hbar4}. In this case a light diquark-heavy quark structure 
for the baryon is favored. 

It is also possible to reach the same conclusion by looking at the mean 
values of the variables $\vec X$ and $\vec Y$. At the dominant order, we have 
\begin{equation}
\label{size}
\left\langle \vec X^{\, 2}\right\rangle=\frac{3+2N_\xi}{a},\quad \left\langle \vec Y^{\, 2}\right\rangle=\frac{3+2N_\eta}{4a}.
\end{equation}
Because of the particular shape of the potential (a Cornell type), the more 
the system is small, the more its mass will be small. Indeed, the energy of 
the flux tubes increases with the size of the baryon, while the attractive 
Coulomb-like forces are larger for small quark separations. 
Equations~(\ref{size}) shows that an excitation of type 
$N_\eta$ will keep the baryon smaller than the corresponding excitation 
in $N_\xi$. Thus, the most favored possibility, at least for the small 
excitation numbers, is also $N_\eta=N$ and $N_\xi=0$. 

As for light baryons, heavy baryons can be labeled by a single harmonic 
oscillator excitation number and the emergence of this quantum number can be 
understood within a relativistic quark model framework. However, we only 
discuss the ground state in the following, that is 
$N_\xi=0$ and $N_\eta=N=0$. Excited states will 
be studied in subsequent papers. 

\subsection{Determination of the parameters}\label{detparam}

The parameters needed for $qqq$ baryons have been obtained in our previous 
papers \cite{lnc,lnc2} but, since we use a new value for $k_0$, we prefer to 
determine a set of new values for the parameters which are gathered in 
Table~\ref{parmod}. The new values are very close to the previous ones and 
do not alter the good results obtained in Refs.~\cite{lnc,lnc2}.
The auxiliary field method 
systematically overestimates the absolute scale of the mass spectrum \cite{Sem03}.
In order to obtain a good accuracy for the baryon masses, it is 
necessary to perform the rescaling $a=\pi\sigma/(6 k_0)$ throughout 
the mass formulas, where $\sigma$ is the physical string tension for a meson \cite{lnc}. 
As $u$ and $d$ current quark masses are expected to be 
very small, we also take a vanishing current mass for the quark $n$. 
The parameter $\sigma$ 
and $f$ are fitted on the $nnn$ baryon Regge trajectory. As it is not possible 
to determine independently $\alpha_0$ and $f$, we choose for $\alpha_0$ a value 
in agreement with other potential models. More details can be found in 
Ref.~\cite{lnc}. 
It is worth noting that the value 3.6 for $f$ is in the range [3-4] and that
the string tension value of 0.165~GeV$^2$ is in good agreement with the value
predicted by the flux tube model \cite{ft95}. 
The $s$-quark mass is fitted to the strange baryon masses in 
the band $N=0$ \cite{lnc2}. 
The value found for $m_s$ is larger than the PDG value of
$104^{+26}_{-34}$~MeV \cite{PDG}. However, a strange quark mass in the
range 0.2-0.3~GeV is quite common in potential models \cite{luch91,bada,naro08b}.

\begin{table}[t]
\centering
\caption{Parameters for $qqq$ baryons.}
\begin{ruledtabular}
\begin{tabular}{lll}
\multicolumn{2}{c}{Fixed parameters} & Fitted parameters \\
\hline
$m_n=0$ & $\alpha_0=0.4$ & $m_s=0.240$~GeV \\
$k_0=0.952$ & $\delta=1.0$~GeV & $\sigma=0.165$~GeV$^2$ \\
$a=\pi \sigma /(6 k_0)$ & $\beta=2.85$ & $f=3.60$ \\
\end{tabular}
\end{ruledtabular}
\label{parmod}
\end{table}

The parameters linked to heavy quarks are $m_c$, $m_b$, $k_1$, and $\alpha_1$. 
We fix $\alpha_1=0.7\alpha_0$ from the quark model study of 
Ref.~\cite{tf_Semay}. The value $k_1=0.930$ has been computed in Ref.~\cite{Bsb04}. 
Because of the rescaling $a=\pi\sigma/(6 k_0)$, only the 
ratio $k_1/k_0\approx 0.98$ is relevant. Let us note that to fix this ratio to 1 does not 
change noticeably the other parameters. The heavy quark masses can be fitted to the 
experimental data as follows. The quark model mass formula~(\ref{M_qqQ}) is 
spin-independent; it should thus be suitable to reproduce the masses of 
heavy baryons for which $J^2_{qq}=0$. Typically, one expects that
\begin{align}
	\left.M_{nnc}\right|_{N=0}&=\Lambda_c = 2286.46\pm0.14\ {\rm MeV}, \\
	\left.M_{nnb}\right|_{N=0}&=\Lambda_b = 5620.2\pm1.6\ {\rm MeV}.
\end{align}
These values are reproduced by formula~(\ref{M_qqQ}) with $m_c=1.252$~GeV and 
$m_b=4.612$~GeV. These masses, obtained by a comparison of the quark model to 
the experimental data, are clearly compatible with those obtained from the 
mass combination~(\ref{E0}) -- both determinations actually differ by less 
than 5\%. This is a first evidence of the compatibility between quark model 
and large $N_c$ expansion in the heavy baryon sector. The supplementary 
parameters for $qqQ$ baryons are gathered in Table~\ref{parmod2}. 
One can notice that
we predict $\left.M_{nsc}\right|_{N=0}=2433$~MeV and 
$\left.M_{nsb}\right|_{N=0}=5767$~MeV with these parameters.
These values are very close to  the experimentally observed masses
of  $\Xi_c$ and $\Xi_b$ respectively.

\begin{table}[ht]
\centering
\caption{Supplementary parameters for $qqQ$ baryons.}
\begin{ruledtabular}
\begin{tabular}{ll}
Fixed parameters & Fitted parameters   \\
\hline
$k_1=0.930$ & $m_c=1.252$~GeV \\
$\alpha_1=0.7 \alpha_0$& $m_b=4.612$~GeV \\
\end{tabular}
\end{ruledtabular}
\label{parmod2}
\end{table}

\section{Comparison of the two approaches}
\label{compar}

First we recall that the heavy quark masses can be obtained by two different 
ways. On the one hand, the large $N_c$ inspired mass combination~(\ref{E0}) 
leads to $m_c=1315$~MeV and $m_b=4642$~MeV. On the other hand, the quark model 
mass formula~(\ref{M_qqQ}) is compatible with the experimental data provided 
$m_c=1252$~MeV and $m_b=4612$~MeV. Both approaches lead to quark masses that 
differ by less than 5\%, as pointed out in Sec.~\ref{detparam}. Thus the two 
approaches that are considered in this paper agree at least at the  
dominant order, where only $m_Q$ is present. 

The other parameter involved in the large $N_c$ mass formula is $\Lambda$. 
A comparison of the spin independent part of the mass formulas (\ref{mlnc}) 
and (\ref{M_qqQ}) leads to the following identification for $N_c = 3$
\begin{eqnarray}  
\label{rel1}
{c_0}&=& \frac{1}{3}\left.M_1\right|_{N=0}\nonumber\\
&=&
\frac{4}{3}\mu_1 - \frac{2}{9}\sqrt{\frac{k_1 a}{3}} (\alpha_0 
+ 2 \sqrt 2\alpha_1 ) - \frac{f a}{3\pi \mu_1},
\end{eqnarray}
where $\mu_1=\sqrt{3k_1a/2}$. 
According to Eqs.~(\ref{largenpar}) and (\ref{lambdadef})
one has $c_0 = \Lambda \simeq 0.324$~GeV.
The quark model parameters of
Tables~\ref{parmod} and \ref{parmod2} give 0.333~GeV for the 
expression after the second equality sign in Eq.~(\ref{rel1}),
which means a very good agreement for the QCD scale $\Lambda$. In this 
quantity, 0.475~GeV comes from the dynamics of the confinement ($4 \mu_1/3$), 
while the Coulomb interaction (term containing $\alpha_0$ and 
$\alpha_1$) contributes with $-0.044$~GeV and the self-energy 
(term proportional to $f$) with $-0.097$~GeV. The mass shift yielded by these 
two residual interactions is quite significant and their presence improves 
the value of $\Lambda$.

Next the terms of order $1/m_Q$ lead to the identity
\begin{eqnarray} 
\label{rel2}
c^{'}_0 &=& 2 m_Q \left.\Delta M_Q\right|_{N=0}\nonumber\\
&=&k_1 a \left[3\left(\sqrt{2}-1\right) \left(1-\frac{f a}{2 \pi \mu_1^2}\right)
-\frac{\alpha_0}{6} \left(\sqrt{2}-1\right) 
+\frac{4 \alpha_1}{3} \right] . 
\end{eqnarray}
Note that to test this 
relation the value of $m_Q$ is not needed, like for the identity~(\ref{rel1}).
The large $N_c$ parameter $\Lambda=0.324$~GeV gives for the left hand side of
(\ref{rel2}) $c^{'}_0\sim \Lambda^2 = 0.096$~GeV$^2$ and the quark model gives 
for the right hand side 0.091 GeV$^2$, which is again a good agreement. In this 
quantity, the contribution of the dynamics of the confinement [$k_1 a F(0)$] 
is 0.105~GeV$^2$, while the 
contributions of the Coulomb interaction and of the self-energy are  
0.029~GeV$^2$ and $-0.043$~GeV$^2$ respectively. The relative magnitude of 
these two terms compared to the first one is here larger but they nearly cancel each 
other. 

The SU(3)-flavor breaking term is proportional to the factor 
$\epsilon\Lambda_\chi \sim m_s - m$ in the $1/N_c$ mass formula (\ref{breaksim}) 
and similarly by the nonvanishing strange quark mass in the quark model 
where one takes $m = 0$. Using Eqs.~(\ref{breaksim}), (\ref{break}), and 
(\ref{M_qqQ}) one obtains 
\begin{eqnarray}
\epsilon\Lambda_\chi&=&\frac{2}{\sqrt 3}\left.\Delta M_{1s}\right|_{N=0}\nonumber\\
&=&\frac{2m_s^2}{\sqrt 3\mu_1} 
\left[ \frac{1}{2} 
- \frac{1}{12 \mu_1}\sqrt{\frac{k_1 a}{3}} \left( \alpha_0 
+ 2 \sqrt 2\alpha_1\right) 
+ \frac{f a}{2 \pi}\left( \frac{3}{4 \mu_1^2}+\frac{\beta}{\delta^2} \right) \right].
\end{eqnarray}
From phenomenology, Eq.~(\ref{break}) implies that   
$\epsilon\Lambda_\chi=0.206$~GeV and the quark model estimate
is $0.170$~GeV, which compares satisfactorily with the value used in the 
combined $1/N_c$ and $1/m_Q$ expansion  \cite{Jenkins:1996de}. In 
the quark model, the contribution 
of the dynamics of the confinement (term proportional to 1/2) is 0.093~GeV, 
while the contributions of the Coulomb interaction and of the self-energy are 
 $-0.009$~GeV and 0.085~GeV respectively. Thus the effect of the 
self-energy is as large as that of the confinement. 

Let us recall that, except for $m_c$ and $m_b$, all the model parameters 
are determined from theoretical arguments combined with phenomenology, or are 
fitted on light baryon masses. The comparison of our results with the 
$1/N_c$ expansion coefficients $c_0$, $c^{'}_0$ and $\epsilon\Lambda_\chi$ are 
independent of the $m_Q$ values. So we can say that this analysis is 
parameter free. 

So far, our formalism is spin-independent. An evaluation of the 
coefficients $c_2$, $c_2'$, and $c_2''$ through a computation of the 
spin-dependent effects within a three-body quark model is then 
\textit{de facto} out of the scope of the present approach. If included, 
the spin-dependent interactions between quarks $i$ and $j$ would appear 
as relativistic corrections to the Coulomb potential, proportional to 
$1/\mu_i\mu_j$ ($\mu_i$ is the dynamical mass of the quark $i$) \cite{lnc}. 
At the dominant order, one expects that $c_2\propto\mu^{-2}_1$ and 
$c_2''\propto\mu^{-1}_1$. The ratio $c_2''/c_2$ should thus be of order 
$\mu_1=356$~MeV, which is roughly in agreement with 
Eq.~(\ref{largenpar}) stating that $c_2''/c_2 \sim \Lambda$. This gives an 
indication that the quark model and the $1/N_c$ expansion method would  
remain compatible if the spin-dependent effects were included, as we 
already pointed out in the light baryon sector \cite{lnc,lnc2}.

Charm and bottom baryons have been studied with a Hamiltonian similar to ours 
in Ref.~\cite{naro08b}. All parameters ($a$, $\alpha_S$, $m_n$, etc.) have
values very close to ours but some differences exist: A genuine junction Y is 
used for the confinement instead of the approximation (\ref{ssh2})-(\ref{pot1}),
the auxiliary fields are introduced only at the level of the 
kinetic part, the Coulomb potential is not treated perturbatively and the
color-magnetic interaction is taken into account. The 
consequences of this procedure is that no analytical mass formula can 
be derived explicitly. But, the numerical results obtained in that paper are 
in good agreement with experiment, which reinforce our approach. Moreover it 
was also found that a unit of angular momentum between the 
heavy quark and the two light quarks is energetically favored with respect to 
a unit of angular momentum between the two light quarks. This correspond to our choice 
$N_\xi=0$.

\section{Conclusions}\label{conclu}

Our previous studies establishing a connection between the quark model and 
the $1/N_c$ expansion for light baryons have been successfully extended to 
baryons containing a heavy quark. Accordingly, the $1/N_c$ expansion was 
supplemented by an $1/m_Q$ expansion due to the heavy quark. As in the light 
baryon sector, there is a clear correspondence between various terms appearing 
in our mass formula~(\ref{M_qqQ}) and those of the mass formula in the 
combined $1/N_c$ and $1/m_Q$ expansion described in Sec.~III. First, both 
methods lead to compatible values for the heavy quark masses. Second, the 
typical QCD scale involved in the $1/N_c$ expansion is well reproduced by 
the quark model without any free parameter: All necessary parameters have 
been previously fitted on light baryons. Finally, the dominant term in 
SU(3)-flavor breaking expansion is satisfactorily reproduced. The 
spin-dependent terms, seen as relativistic 
effects, deserve a special study, to be considered in the future.

This study, completing the two previous ones \cite{lnc,lnc2}, brings reliable 
QCD-based support in favor of the constituent quark model assumptions due to the 
compatibility of its mass formula and the mass formula derived from the model 
independent $1/N_c$ expansion. 
Moreover, better insight into the coefficients $c_i$ encoding the QCD dynamics 
in the mass operator is obtained: the dependence on the quark content and on 
the excitation number.

We presently focused on ground state heavy baryons. For excited states, 
the quark model suggests that the band number $N$ classifying the heavy 
baryon resonances should be associated to the quantum of excitation of 
the heavy quark--light diquark pair in a harmonic oscillator picture. We 
leave a detailed study of excited heavy baryons for future studies.

\acknowledgments
Financial support is acknowledged by C.S. and F.B. from F.R.S.-FNRS (Belgium).

\end{document}